\documentclass[usenatbib,onecolumn]{mn2e}
\usepackage{txfonts}
\usepackage{epsfig}
\usepackage{lscape}
\usepackage{graphicx}

\def\newline{\hfil\break}

\title[Galaxy Formation in WDM Cosmology]{Galaxy Formation in Warm Dark Matter Cosmology}
\author[Menci et al.]{N. Menci, F. Fiore, A. Lamastra\\
INAF - Osservatorio Astronomico di Roma, via di Frascati
33, I-00040 Monteporzio, Italy\\
}
\pagerange{\pageref{firstpage}--\pageref{lastpage}} \pubyear{}

\begin{document}
\maketitle
\vspace{1cm}
\begin{abstract}
We investigate for the first time the effects of a Warm Dark Matter (WDM) power spectrum on the statistical properties of galaxies using a  semi-analytic model of galaxy formation. The WDM spectrum we adopt as a reference case is suppressed - compared to the standard Cold Dark Matter (CDM) case - below a cut-off scale $\approx1$ Mpc corresponding (for thermal relic WDM particles) to a mass $m_X=0.75$ keV. This ensures consistency with present bounds provided by the microwave background WMAP data and by the comparison of hydrodynamical N-body simulations with observed Lyman-$\alpha$ forest. We run our fiducial semi-analytic model with such a WDM  spectrum to derive galaxy luminosity functions (in B, UV, and K bands) and the stellar mass distributions over a wide range of cosmic epochs, to compare with recent observations and with the results in the CDM case. The predicted color distribution of galaxies in the WDM model is also checked against the data. When compared with the standard CDM case, the luminosity and stellar mass distributions we obtain assuming a WDM  spectrum are characterized by: i) a flattening of the faint end slope and ii) a sharpening of the cutoff at the bright end for $z\lesssim 0.8$. We discuss how the former result is directly related to the smaller number of low-mass haloes collapsing in the WDM scenario, while the latter is related to the smaller 
number of satellite galaxies accumulating in massive haloes at low redshift, thus suppressing the accretion of small lumps on the 
central, massive galaxies. These results shows how a adopting a WDM power spectrum may contribute to solve two major problems of CDM galaxy formation scenarios, namely, the excess  of predicted faint (low mass) galaxies at low and - most of all - high redshifts,  and the excess of bright (massive) galaxies at low redshifts.
\end{abstract}


\section{Introduction}

The present understanding of galaxy formation in a cosmological context is based 
upon the collapse via gravitational instability of tiny Dark Matter (DM) perturbations 
dominating the matter contribution $\Omega_M\approx 0.3$ to the energy density of the Universe,
with baryons contributing only a fraction $\Omega_b\approx 0.04$. In such a context, the number density 
and the properties of galaxies with different mass ultimately arise from the power spectrum
of DM perturbations at the epoch of recombination. For example, in a Universe dominated by 
Cold Dark Matter (CDM) the low thermal speed of DM particles makes the density perturbations gravitationally unstable down to mass scales negligible for galaxy formation, and structure formation proceeds bottom-up with the progressive collapse of larger and larger 
regions; massive galaxies, groups and clusters form by subsequent aggregations. When coupled with the evidences for a positive 
vacuum energy $\Lambda$  dominating the energy-density of the Universe (with $\Omega_{\Lambda}\approx 0.7$ 
see Spergel et al. 2007), such a $\Lambda$CDM scenario appears to be consistent with a wide range of observables; 
from the temperature anisotropies  in the microwave background radiation (Komatsu et al. 2011) to the clustering 
of galaxies on large scales (Cole et al. 2005). 

However, such observables directly probe scales from$\sim 1$ Gpc down to $\sim 10$ Mpc. At galactic and sub-galactic scales $\lesssim 
1$ Mpc, the computations based on N-body simulations pose several problems to the $\Lambda$CDM scenario. For example, the CDM haloes are predicted to have inner density profiles much steeper than those inferred from the rotation curves of real galaxies (Moore et al. 1999, Abadi et al. 2003, Reed et al. 2005; Madau, Diemand, Kuhlen 2008; for observations see, e.g., Gentile et al. 2007, McGaugh et al. 2007), the discrepancy being larger for dwarf galaxies (see Governato et al. 2007; Abadi et al. 2003; de Blok et al. 2008). In addition, CDM haloes are predicted to contain a wealth of substructure, which we expect to observe as satellite galaxies within the haloes of galaxies and groups, in contrast with observations (see Klypin et al. 1999, Moore et al. 1999; see also Mateo et al. 1998). 

Since at such small scales the model predictions depend on the complex physics involved in the process of galaxy formation, part of the 
solutions to the above problems may be of  astrophysical nature. Feedback processes such as heating or winds caused by Supernovae explosions and heating due to the UV background may be effective in  suppressing star formation in the low-mass haloes of satellite galaxies (Bullock et al. 2000; Benson et al. 2002; Somerville et al. 2002; Kravtsov, Gnedin \& Klypin 2004;  Governato et al. 2007; Mashchenko, Wadsley \&  Couchman 2008), and in flattening the the inner density profiles in both spirals and dwarfs (see Governato et al. 2007; 2010). 

However, the origin of the above problems may also lie, at least partially, in the shape of the CDM power spectrum, which could 
over predict the amplitude of small-scale ($\lesssim 1$ Mpc) perturbations. Indeed, a strong indication that this may be the 
case comes from the statistical properties of galaxies and Active Galactic Nuclei (AGN), usually investigated in the framework of  
$\Lambda$CDM  through semi-analytic models of galaxy formation  (SAMs, see Kauffman,  White \& 
Guiderdoni 1993, Cole et al. 1994; Monaco, Salucci \& Danese 2000, Kauffmann \& Haenhelt 2000; Granato et al. 2004; Menci et al. 2006; Croton et al. 2006; Bower et al. 2006; Marulli et al. 2008), which connect the physical processes involving gas and star formation to the merging histories (and hence to the power spectrum) 
of DM haloes. All such models (including different astrophysical feedback process) 
concur in predicting an excess of small-mass galaxies compared to observations. In particular:  i) the predicted galaxy luminosity functions at low redshift are much steeper than observed 
unless a strong feedback is assumed (see, e.g., Somerville, Primack 1999; Cole et al. 2000; Menci et al. 2002), while at high redshifts
the larger escape velocities of galaxies (due to their larger densities) make the feedback inefficient in suppressing star formation 
in low-mass objects. This results into an over-prediction of faint galaxies increasing with redshift  when compared, e.g., with the evolution of the K-band luminosity function for $0\lesssim z\lesssim 3$  (see Cirasuolo et al. 2010) or with the faint Lyman-break galaxies (Lo Faro et al. 2009) at $z\gtrsim 3$; 
the excess of star-forming low-mass galaxies at $z\gtrsim 3$ reflects into an excess of red low-mass galaxies at $z\approx 0$ 
(see, e.g., Croton et al. 2006; Salimbeni et al. 2008). 
ii) The number of galaxies with stellar mass $M_*\lesssim 10^{10}$ M$_{\odot}$  in the Universe is systematically over predicted by
all theoretical CDM models (Fontana et al. 2006; Fontanot et al. 2009; Marchesini et al. 2009) in the whole redshift range $z\lesssim 3$ where the small-mass end of the mass function has been measured (see also Santini et al. 2012).

Thus, the baryonic processes (and the feedback in particular) implemented so far in all galaxy formation models do not seem able to  provide an explanation for the {\it predicted} excess. While it is still possible that the problem stems from a poor understanding of 
the laws governing star formation especially at high redshifts $z\gtrsim 2$ (see, e.g.,  Krumholz \& Dekel 2011), 
it is interesting to note that the over-predictions of low-mass objects in the CDM scenario concerns also the AGN population; 
recent estimates of the X-ray luminosity functions of AGN extending down to log $L$(2 -10 keV)$\approx 42.75$ (Fiore et al. 2011) 
are in sharp contrast with the large density of faint AGN predicted by analytic (Shankar 2009) or semi-analytic (Menci et al. 2006, 2008) models  for the co-evolution of galaxies and AGNs (se also Shankar \& Mathur 2007; Wyithe \& Loeb 2003; Lapi et al. 2006, Shen 2009; Shankar 2010 and references therein), even when the observed luminosity functions are corrected for the estimated fraction of obscured objects (e.g., La Franca et al. 2005). 

Although, individually, the above problems do not definitely rule out the CDM model due to the complexity of the baryonic physics relating DM haloes with observable properties, the concurring evidences listed above call for a more radical approach, to investigate the possibility that the conflicts are rooted in an 
excess of power of the CDM spectrum at small scales, and that density perturbations could be dumped below same characteristic scale $r_{fs}$. A natural mechanism for such  a suppression is the free-streaming of DM particles
on scales $r_{fs}\sim v_s\,t_{eq}$, where $v_s$ is the DM thermal speed at the time of matter-radiation equality $t_{eq}$. For large 
DM velocities $v_s/c\approx 0.2$ (corresponding to a neutrino with mass $\approx 12 $ eV) this leads to the well-known Hot DM scenario (for a review see Primack \& Gross 2000) which would yield $r_{fs}\approx 25$ Mpc, well above the most recent observational limits $r_{fs}\lesssim 0.1-1$ set by the comparison of the Lyman-$\alpha$ forest spectra at $z\approx 2.5$ with the N-body simulations (Viel et al. 2005; 2008). However, for DM particles with larger mass $m_X$ the streaming speed could be reduced so as to bring $r_{fs}$ close
to $1$ Mpc (corresponding to a mass scale $M_{fs}\approx 10^{9\div 10}\,M_{\odot}$), consistent with present bounds but still large enough to affect galaxy formation. If constituted by thermal relic particles, such Warm DM (WDM) would have streaming velocities $v_s\propto T_X/m_X$, where $T_X$ is the effective temperature determining  the value of the WDM density $\rho_X\propto m_X\,T_X^3$, so that $r_f\propto \Omega_X^{1/3}\,m_X^{-4/3}$. Thus, to lower the smoothing scale $r_f$ of a factor $\approx 10^2$ compared to the above Hot DM case, the WDM particles should have  $m_X\sim 1$ keV if their density has to match the observed DM density parameter $\Omega_X\approx 0.25$. This type of relic WDM was discussed by Peebles (1982), Bond, Szalay \& Turner (1982), and could have a physical counterpart in the gravitino, the supersymmetric partner of the graviton, in  gauge-mediated supersymmetry breaking models (Pagels, \& Primack 1982; see Steffen 2006 for a recent review). Another possibility is that WDM particles have never been in thermal equilibrium, as is the case, e.g., for sterile neutrinos created from oscillations with active neutrinos (Olive \& Turner 1982; Dodelson \& Widrow 1994; Shi \& Fuller 1999; Abazajian, Fuller \& Patel,  2001; Dolgov \& Hansen 2002 ). A discussion on the WDM candidates can be found in Colombi, Dodelson, Lawrence (1996, and references therein). For scenarios based on composite DM  see Khlopov et al. (2008).
 
Previous studies based on N-body simulations have shown that WDM can resolve some of the tension between theoretical predictions and observations (Bode et al. 2001; Avila-Reese et al. 2001). Some simulations suggest that the density profiles 
of WDM haloes are shallower in the inner regions (Colin,  Valenzuela, Avila-Reese 2008; see also de Vega, Salucci, Sanchez 2010), and that WDM can provide an explanation for the observed kinematics of satellites of massive galaxies (Lovell et al. 2011). Simulations focussed on the halo properties and mass distribution in WDM scenario (see Knebe et al. 2002; Zavala et al. 2009; Polisensky \& Ricotti 2011, Smith \& Markovic 2011) have shown that there is a sensible suppression in the number of haloes with $M<M_{fs}$. However, the effects of a WDM spectrum on the properties of galaxies at such mass scales would be best investigated with SAMs, since not only they allow to probe in detail the effects on the halo statistics for  $M<M_{fs}$, but also they can investigate the chain of effects that such a suppression has on the baryonic processes involved in galaxy formation: a modification of the power spectrum changes  the number and the formation epoch (and hence the internal densities) of the of high-redshift progenitors of local galaxies, an effect which propagates to gas cooling, to the disk sizes, and to star formation; this also affects both dynamical friction and binary merging, the processes determining the fate of sub-haloes associated to satellite galaxies (see, e.g., Somerville \& Primack 1999; Menci 2002), and consequently the stellar mass growth of large galaxies due to accretion of smaller objects; finally, starbursts and AGN accretion and feedback (included in  almost all state-of-the-art SAMs) induced by galaxy interactions are affected by the change in the number and in the properties of sub-haloes. 

In the present paper we investigate for the first time the effects of a WDM power spectrum on the statistical properties of galaxy using a  SAM of galaxy formation. We adopt a WDM power spectrum with a power suppression (compared to the CDM spectrum) below  a cut-off scale $\approx 1$ Mpc, to retain consistency with present bounds provided by the microwave background WMAP data and by the comparison of hydrodynamical N-body simulations with observed Lyman-$\alpha$ forest (see Viel et al. 2005). 
In terms of the properties of the WDM particles, such a cut off scale $\approx 1$ Mpc is achieved for a mass of DM particles $m_X\sim 1$ keV (see Sect. 2) whose 
free streaming length is $r_{fs}\approx 0.2$ Mpc, corresponding to the wavelength of the mode for which the linear perturbation amplitude is suppressed by a factor of 2.  
We run our fiducial semi-analytic model (Menci et al. 2006, 2008) with such a WDM 
power spectrum to derive galaxy luminosity functions (in B, UV, and K bands) and the stellar mass distributions over a wide range of cosmic epochs, to compare with recent observations (Sect. 3); the predicted color distribution of galaxies in the WDM model is also checked against the data. Our results are discussed in Sect. 4, while Sect. 5 is devoted to conclusions.
\section{Method}
The SAM we use connects the physical processes involving baryons 
(physics of gas, star formation, feedback, growth of supermassive Black Holes) to the merging histories
of DM haloes, ultimately determined by the DM initial power spectrum. In the present paper we use the 
same model developed by Menci et al. (2005, 2006, 2008) with the same choice of free parameters; the only change 
is that merging trees are computed in a WDM cosmology. This allows to single out the effects of changing the DM spectrum with the same 
baryon physics. Note that in the present paper the tree resolution as been increased with respect to Menci et al. (2005) to resolve 
DM masses down to $10^7$ M$_{\odot}$ for both the WDM and the CDM case, to better investigate the effect of changing the spectrum on the low-mass end of the mass distribution.

Here we briefly recall the basic features of the model (for a complete description see Menci et al. 2005, 2006, 2008), 
and we describe the power spectrum we use to compute the merging trees in the WDM cosmology. 
\subsection{The Semi-Analytic Model}

Galaxy formation and evolution is driven by the collapse and growth of DM haloes, which originate 
from the gravitational instability of  overdense
regions in the primordial density field. This is taken to be a random,
Gaussian  density field within the
''concordance cosmology" (Spergel et al. 2007), for which we adopt round
parameters  $\Omega_{\Lambda}=0.7$, total matter density parameter $\Omega_{M}=0.3$ 
(with baryons contribution corresponding to  $\Omega_b=0.04$ and Hubble constant (in units of 100 km/s/Mpc) $h=0.7$. The
normalization of the spectrum is taken to be $\sigma_8=0.9$ in terms of the variance
of the field smoothed over regions of 8 $h^{-1}$ Mpc. Adopting the exact best fit 
WMAP7 values for the above parameters does not change our results appreciably. 

As  cosmic time increases, larger and larger regions of the density field
collapse, and  eventually lead to the formation of groups and clusters of
galaxies; previously formed, galactic size  condensations are enclosed. 
The corresponding merging rates
of the DM haloes are provided by the Extended Press \& Schechter formalism (EPS, 
see Bond et al. 1991; Lacey \& Cole 1993); this depends on the variance 
of the power spectrum on a mass scale $M$
\begin{equation} 
\sigma^2(M)=\int {dk\,k^2\over 2\,\pi^2}\,P(k)\,W(kr)
\end{equation}
where $P(k)$ is the linear power spectrum of DM perturbations at a wavelengths 
$k=2\pi/r$, $M\simeq 1.2\,10^{12}\,h^2\,{\rm M}_{\odot}\,(r/{\rm Mpc})^3$ is the total mass 
in a sphere of radius $r$, and $W$ is the top-hat window function (see 
Peebles 1993).

The clumps included into larger DM haloes
may survive as satellites, or merge to form larger galaxies due to binary
aggregations,  or coalesce into the central dominant galaxy due to dynamical
friction (see Menci et al. 2005, 2006); in addition satellite halos are partially disrupted 
as the density in their outer parts becomes 
lower than the density of the host halo within the pericentre 
of its orbit (see Menci et al. 2002 for details). Since the above galaxy coalescence 
processes take place over time scales that grow longer over
cosmic time,  the number of satellite galaxies increases as the DM host haloes
grow from groups to clusters. All the above processes are implemented in our
model  through a Monte Carlo realization of merging trees (generated with EPS probabilities 
depending on the spectrum through eq. 1) and of the dynamical processes affecting 
the sub-haloes once they are included into larger haloes, following the canonical
prescriptions of SAMs. 

We do not expect our results to be 
appreciably affected by uncertainties in the description  of the above processes. 
Although some variance exists between the galaxy merging times predicted by N-body simulations and the
different  SAMs (see De Lucia et al. 2010 for a discussion), there 
is an overall agreement in terms of the mass build up. Indeed, 
Henriques et al. (2011) compare the evolution of the K-band luminosity distribution predicted by three different SAMs,  
including the one used here and the De Lucia \& Blaizot (2007) model.
The latter model, which adopts galaxy merging histories extracted from high-resolution N-body simulations, 
differs at most by a factor two in the luminosity of the brightest 
galaxies from the present SAM; since the major fraction of such effect is due to the inclusion 
of a Radio Mode AGN feedback in the De Lucia \& Blaizot (2007) model (contributing to suppress
the K-band luminosities by about 1 magnitude, see Croton et al. 2006), 
the difference in the growth of massive galaxies due to uncertainties in the 
treatment of dynamical processes of sub-haloes is negligible. Similar conclusions
have been presented by Fontanot et al. (2009).

While the above dynamical sector of the model is entirely determined by the initial power spectrum, 
the observable properties of galaxies are determined 
by the physics of baryons, which is related to the 
DM merging trees through the canonical approach of SAMs.
The radiative gas cooling, the ensuing star formation and the
Supernova events with the associated feedback occurring  in the growing
DM haloes are computed for each sub-halo hosting the galaxy. 
The cooled gas with mass $m_c$ settles into a rotationally supported disk with radius $r_d$
(typically ranging from $1$ to $5 $ kpc), rotation velocity $v_d$
and dynamical time $t_d=r_d/v_d$, all related to the DM sub-halo mass. The gas gradually condenses  into
stars at a rate $\dot m_*\propto m_c/t_d$; the stellar ensuing feedback
returns part of the cooled gas to the hot gas phase
at the virial temperature of the halo. An additional
channel for star formation implemented in the model is provided by
interaction-driven starbursts, triggered not only by merging but
also by fly-by events between galaxies; such a star formation mode
provides an important contribution to the early formation of stars
in massive galaxies, as described in detail in Menci et al. (2003,
2005). 

The model also describes the growth of Black Holes (BHs) inside each DM halo from primordial seeds 
with mass $M_{BH}\approx 10^2\,M_{\odot}$ through merging between galaxies and 
accretion of a fraction  of the cold galactic gas $m_c$. The accretion is triggered by 
galaxy encounters, and the fraction of cold gas funnelled to the central BH 
is derived from the model by Cavaliere \& Vittorini (2000). The active accretion phase of BHs
corresponds to the AGN, which release part of their radiation energy in the 
surrounding interstellar medium, thus heating and expelling part of the galactic gas 
(Lapi, Cavaliere, Menci 2005); the implementation of such a form of AGN feedback in the 
semi-analytic model is described in detail in Menci et al. (2008). 

\subsection{Implementing the WDM power spectrum}

Detailed calculations of the free-streaming scale $r_{fs}$ yield (see Smith \& Markovic and references therein)
\begin{equation}
r_{fs}\approx 0.2\Big[{\Omega_X\,h^2\over 0.15}\Big]^{1/3}\,\Big[{m_X\over {\rm keV}}\Big]^{-4/3}\,{\rm Mpc}
\end{equation}
The lighter is the DM particle  the larger is the free-streaming length $r_{fs}$, corresponding 
to the scale below which the perturbations are wiped out. 
In the case WDM is composed by relic thermalized particles, the suppression (with respect to the CDM case) of density perturbations in WDM scenarios is quantified 
through the ratio of the two linear power spectra (transfer function) which can be parametrized as (Bode, Ostriker \& Turok 2001)
\begin{equation}
{P_{WDM}(k)\over P_{CDM}(k)}=\Big[1+(\alpha\,k)^{2\,\mu}\Big]^{-5\,\mu}, 
\end{equation}
where the CDM power spectrum $P_{CDM}(k)$ is computed following Bardeen et al. (1986). 
Fitting integrations of the full Einstein-Boltzmann integration (see Viel et al. 2005)  yields $\mu=1.12$ and 
\begin{equation} 
\alpha=0.049 \,
\Big[{\Omega_X\over 0.25}\Big]^{0.11}\,
\Big[{m_X\over {\rm keV}}\Big]^{-1.11}\,
\Big[{h\over 0.7}\Big]^{1.22}\,h^{-1}\,{\rm Mpc}. 
\end{equation}
The above relation between transfer function and DM particle mass $m_X$ is valid for thermal relics; a similar relation 
holds for sterile neutrinos provided one substitutes the mass $m_X$ with a mass $m_{sterile}=4.43\,{\rm KeV}\,
(m_X/{\rm keV})^{4/3}\,(\Omega_{WDM}\,h^2/0.1225)^{-1/3}$ (but see Kusenko 2009 for variations due to their mode of production). In both cases, 
the smaller the WDM mass $m_X$ (or $m_{sterile}$) the larger the suppression with respect to the standard CDM spectrum. 
However, observational upper limits on the power suppression (corresponding to lower limits on $m_X$) have been derived 
by Viel et al. (2005) by comparing the observed Lyman-$\alpha$ forest in absorption spectra of Quasars at $z=2-3$ 
with the results of N-body simulations run assuming different WDM  power spectra of perturbations. As a result, for thermal 
relics a lower limit $m_X\gtrsim 0.6$ keV has been obtained, corresponding to $m_{sterile}\gtrsim 2.5$ keV. 
Since our aim is to estimate the possible impact of WDM on galaxy formation, we shall assume the largest  
suppression of the DM power spectrum which is still safely above the constraints set by the Lyman-$\alpha$ forest; 
thus we adopt as a reference case a  WDM particle mass $m_X=0.75$ keV, and compute the transfer function 
after eq. (3). The corresponding power spectrum, 
which constitutes the basic input for the merging trees at the basis of our semi-analytic model, is shown in Fig. 1. and 
compared with the standard CDM case. 

The power spectrum determines the merging histories and the mass distribution of DM haloes as follows; starting from a grid of parent DM halo masses at redshift $z=0$
(whose abundance is determined from the standard Press \& Schechter 1974 mass function), we fragment them 
into progenitor masses at earlier cosmic times through a Monte Carlo simulation, with fragmentation probabilities 
given by the EPS theory (again dependent on the power spectrum through the variance in eq. 1). By construction, the EPS 
probabilities lead to merging trees where -  at each time level  - the average number of generated DM haloes of given mass 
reproduces the Press \& Schechter mass distribution at the corresponding cosmic time.
Variants in this process associated with a modified Press \& Schechter approach are discussed in Sec. 4.
On the basis of such DM merging trees, we run our SAM. We start by populating with baryons 
the DM haloes at the highest redshifts ($z=10$) in each DM merging tree; then we compute all the physical processes involving baryons as 
described in Sect. 2.1, thus associating a galaxy (with disk gas, stars and a star formation rate) to each DM halo; when the DM haloes merge, 
the galaxies associated with the progenitor DM haloes may coalesce into a larger galaxy or survive as satellites 
according to the processes described in Sect. 2.1. 

\begin{center}
\vspace{-0.cm}
\scalebox{0.55}[0.55]{\rotatebox{-90}{\includegraphics{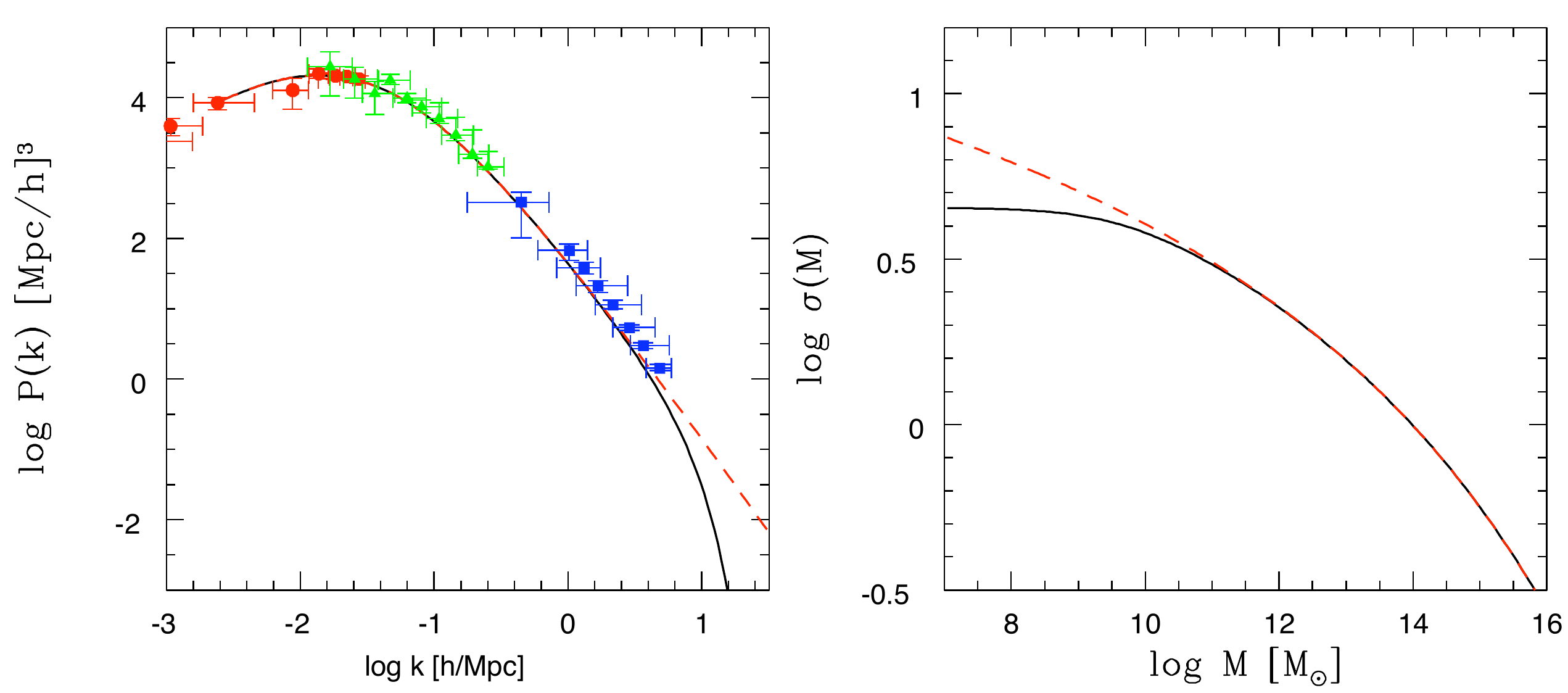}}}
\end{center}
\vspace{-0.1cm }
 {\footnotesize 
Fig. 1. - Left Panel: The linear power spectrum computed for 0.75 keV WDM  particles after eqs. 3 and 4 (solid line) is compared with the CDM spectrum (dashed line) and with different data (see Tegmark, Zaldarriaga 2002 and references therein) derived from fluctuations of the microwave background radiation (WMAP measurements, circles), galaxy clustering (triangles) and Ly-$\alpha$ forest (squares).
\newline
Right Panel: The rms amplitude of perturbations as a function of the mass scale corresponding to the power spectra in the left panel: 
solid line refers to the WDM case, while dashed line to CDM. 
\vspace{-0.1cm}}
\vspace{-0.4cm}
\section{Results}

We now proceed to investigate the effect of implementing the WDM  power spectrum on galaxy formation. We show in Fig. 2 the 
low redshift B-band galaxy luminosity function,  and its UV counterpart at high redshift $z=4$; the luminosity functions in WDM 
cosmology are compared with data and with the corresponding CDM case. 

A first effect of adopting the WDM spectrum is  a decrease in the number of faint objects at both low and high redshifts. This is expected due to the suppression in the amplitude of initial perturbations 
(see Fig. 1). Quantitatively, the suppression of a  factor $\sim 4$ for $M_{b_j}\gtrsim -15$ with respect to the CDM case is similar to that obtained for the halo mass function by Smith \& Markovic (2011) for DM halo masses $M\approx 10^{10}$ M$_{\odot}$ for the same value of $m_X$, and directly derives from the smaller number of low mass DM halos that collapsed at high redshift and survived to form faint galaxies. However, a second interesting effect of adopting a WDM power spectrum appears at low redshifts (Fig. 2, left panel);
at bright magnitudes, the luminosity function shows a steeper cut off compared with the CDM case. This is due to the fact that in large DM haloes at low redshift the growth of massive galaxies due to the accretion of low mass objects is suppressed due to the smaller number 
of satellite galaxies which accumulate in the host DM halo.
\begin{center}
\vspace{0.cm}
\scalebox{0.5}[0.5]{\rotatebox{0}{\includegraphics{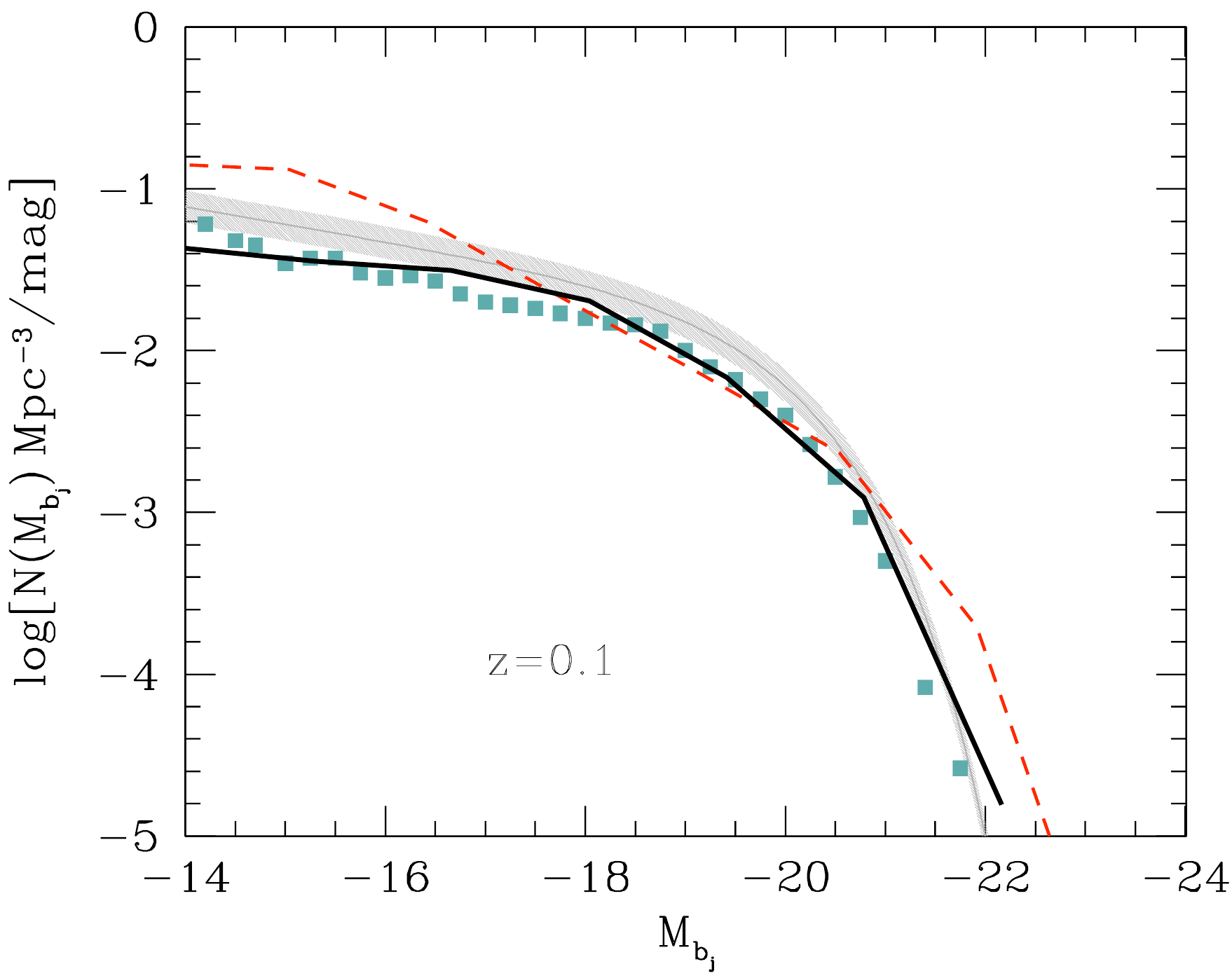}}}
\scalebox{0.5}[0.5]{\rotatebox{0}{\includegraphics{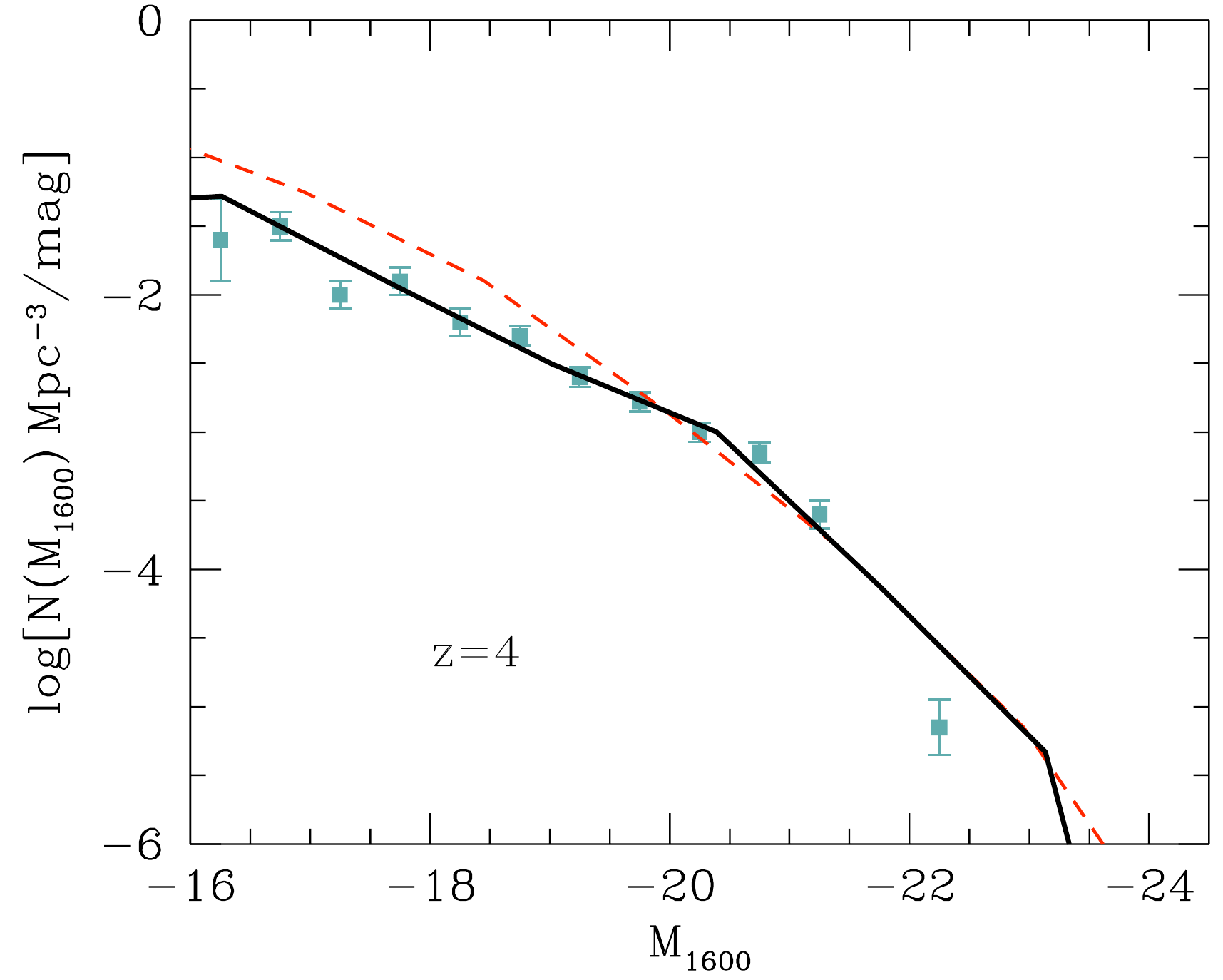}}}
\end{center}
\vspace{-0.3cm }
 {\footnotesize 
Fig. 2. - Left Panel: The local $b_j$ galaxy luminosity function in the WDM model (solid line) is compared with the standard CDM case (dashed line) and with data from the 2dF  (Madgwick et al. 2002, squares) and Sloan (Blanton et al. 2001, hatched region) surveys. Right Panel: The predicted UV luminosity function of drop out galaxies at $z=4$ (solid line WDM, dashed lilne CDM) is compared with data 
from Bouwens et al.(2007).
\vspace{0.2cm}}

Thus, changing the power spectrum in a way consistent with present observational limits seems not only to provide a viable solution for the over prediction of low-mass objects typical of CDM cosmology, but also to contribute to solve a long-standing problem of CDM  
models, namely, the over prediction of bright galaxies at low redshifts. The latter problem has been only recently alleviated by the inclusion of the Radio Mode feedback (see, e.g., Bower et al. 2006, Cattaneo et al. 2006; Croton et al. 2006; Kang, Jing, Silk 2006);  however, at present such a feedback, associated with a low-accretion state of BHs in massive galaxies (presumably at the origin the radio activity), does not 
have a clear observational counterpart, since the observed number of radio galaxies is much smaller than that predicted by SAMs (which need to associate such a Radio Mode accretion to
each massive galaxy with stellar mass $M_*\geq 10^{11}$ $M_{\odot}$, see Fontanot et al. 2011); in addition, at present the implementation of a Radio Mode  still leads to predict a blue fraction of central galaxies that is too high and with an inverted luminosity dependence compared with what observed (see Weinmann et al. 2006).

\begin{center}
\vspace{0.2cm}
\scalebox{0.6}[0.6]{\rotatebox{0}{\includegraphics{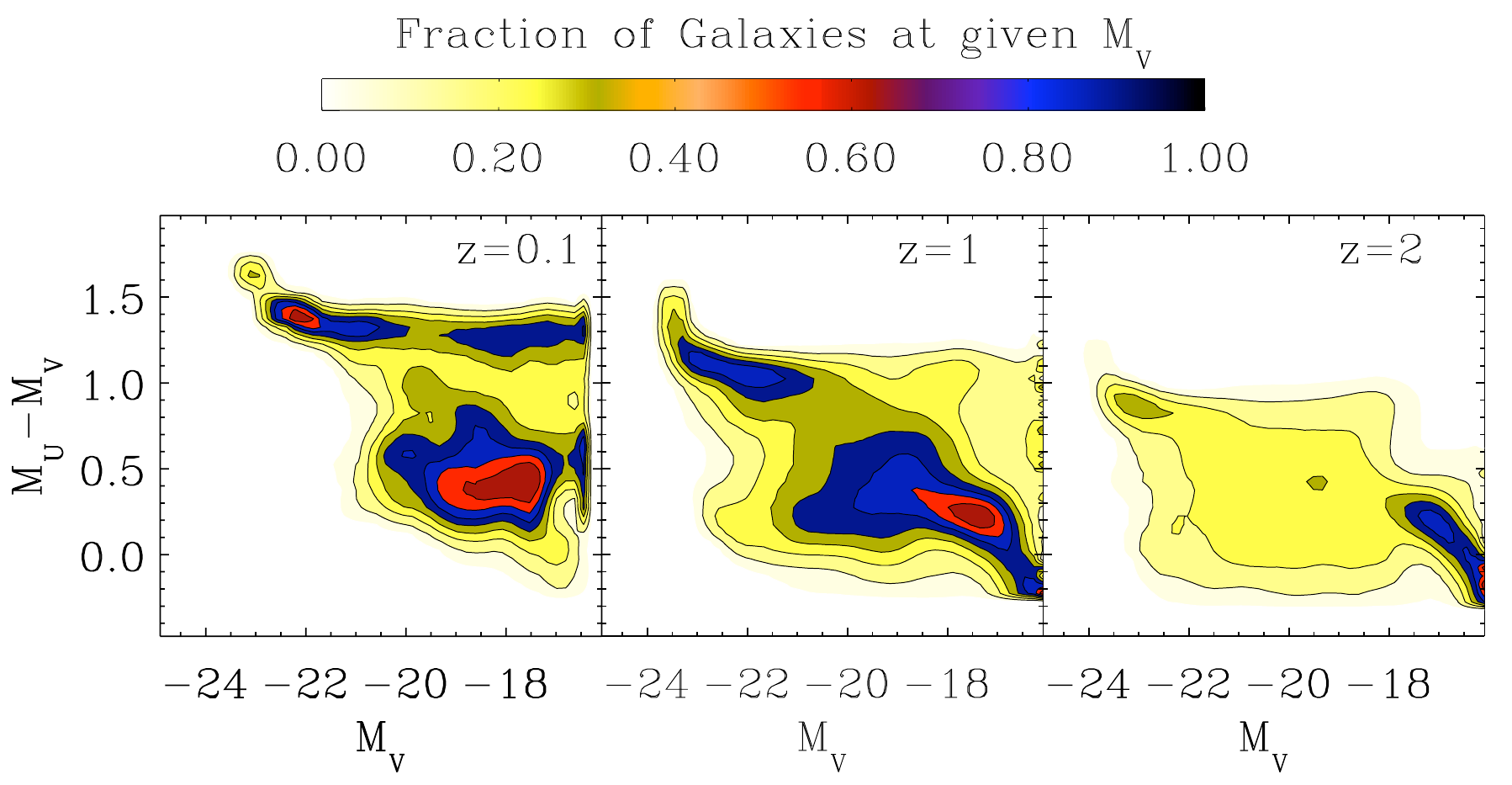}}}
\end{center}
\begin{center}
\vspace{-0.3cm}
\scalebox{0.5}[0.5]{\rotatebox{-90}{\includegraphics{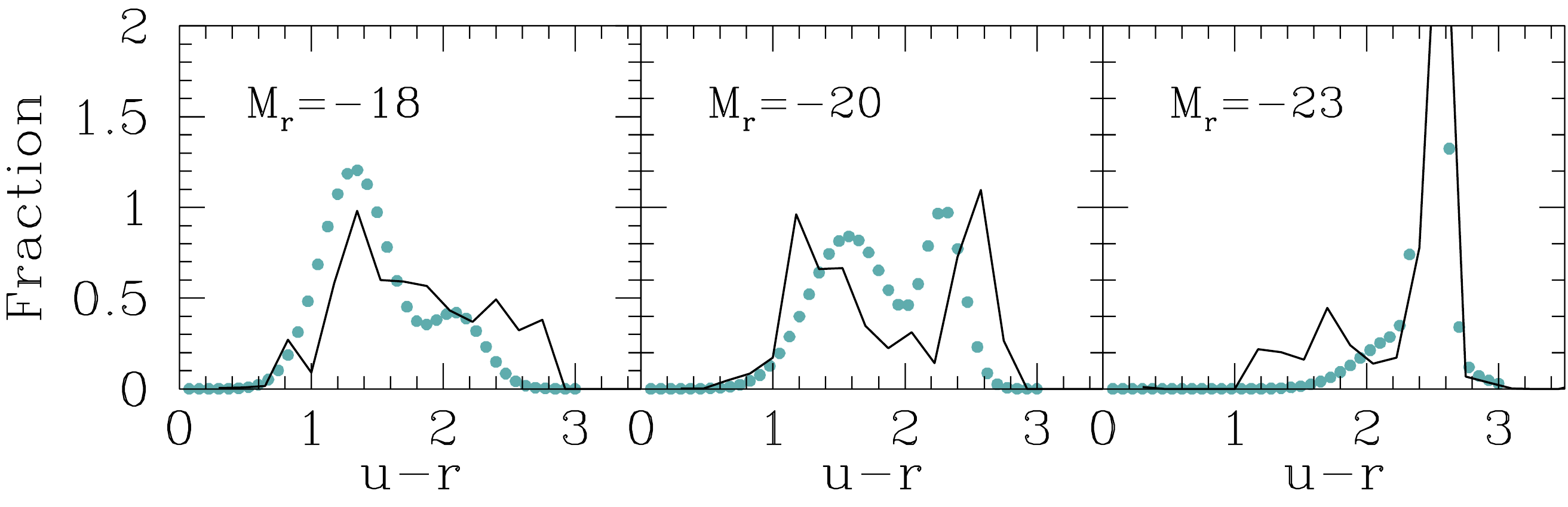}}}
\end{center}
\vspace{0.2cm }
 {\footnotesize 
Fig. 3. - Top Panel: The predicted color distributions for the WDM model at $z=0.1$, 1, 2. The color code represents the fraction of galaxies that, for a given absolute magnitude, are found in different color bins.  
Bottom Panel: Predicted rest-frame $u-r$ color distributions for the WDM case are 
compared with the  Gaussian fit to  the data from the Sloan survey (from Baldry et al. 2004, stars) for different magnitude bins.}
\vspace{0.1cm}

Note that the proposed approach leaves unchanged the baryonic mechanisms affecting the faint end (the Supernovae feedback) and the bright end (the AGN feedback) of the galaxy luminosity distribution. Indeed, we have checked that adopting a WDM spectrum leaves unchanged the  properties of the predicted color-magnitude diagram, which is still characterized by a bimodal distribution in agreement with observations (see Strateva et al. 2001; Baldry et al. 2004) which start to appear already at $z\approx 1.5-2$ as observed (see Bell et al. 2004, Giallongo et al. 2005); a quantitative comparison with detailed observed color distribution for different magnitudes (Fig. 3 bottom panel)  shows an overall agreement with existing measurements , although a moderate excess of bright ($M_r=-23$) galaxies with $u-r\leq 2$ is still present. Note that such excess would be larger (to include a fraction $\sim 1/3$ of bright galaxies) in the absence of AGN feedback.

To investigate how the two effects of adopting WDM power spectrum evolve with redshift, and thus to establish how the different number of satellites in common DM haloes affect the mass growth of galaxies in the WDM vs. the CDM case, we compare with statistical observables more directly related to the stellar mass. The  evolution of the K-band luminosity function in the WDM  case is compared with that derived for the CDM cosmology and with the observational data in the top panel of Fig. 4.; the same comparison is 
performed for the evolution of the stellar mass function in the bottom panel. 

These two observables are related in the model since the first is computed by convolving the stellar Spectral Energy Distributions (SEDs) with the star formation history of the galaxy progenitors, while the latter is the time integral over the same history. The stochastic nature of the merging and star formation histories results in a whole distribution of computed galaxies in the $M_K-M_*$ plane that 
we have checked to be consisted with the observed distribution. On the observational side, the two quantities are measured in different ways. 
The  K-band luminosity functions are taken from the Ultra Deep Survey (UDS), the deepest survey from the UKIRT Infra-Red Deep Sky Survey (UKIDSS), containing imaging in the J - and K-bands, with deep multi-wavelength coverage in $B$ $V$ $R$ $i$' $z$'  
filters in most of the field. The sample contains ≈ 50,000 objects with high completeness down to $K \leq 23$ (Cirasuolo et al. 2010). As 
for the observed stellar mass functions, we compare mainly with the data from Santini et al. (2012);  stellar masses were estimated by fitting  a 14 bands photometry 
(up to 5.5 $\mu$m rest-frame) to the Bruzual \& Charlot synthetic models of stellar populations; the analysis of observations taken with the Hubble Wide Field Camera 3 in the GOODS-S  field allows to achieve an excellent determination of the low-mass-end of the distribution down to small stellar masses $M_{*}\approx 7.6\,10^9$ M$_{\odot}$ even at the highest redshifts ($z\approx 3$). 
Note that, while the stellar mass function  constitutes a direct probe of the effect of WDM spectrum on the growth of stellar mass of galaxies, observational determinations of such a quantity are prone to several observational uncertainties connected, e.g., to the estimate of metallicities or extinction curves necessary to derive  the stellar masses, to the treatment of the TP-AGB phase, or to the reconstruction of the star formation history of each galaxy, that is 
necessary to estimate the appropriate $M/L$ ratio and that may be poorly described by simplistic models like those adopted in 
the stellar synthesis codes (Maraston et al. 2010; Lee et al. 2010). Although the above systematic uncertainties (not included in the error bars in the bottom panel of fig. 4) do not allow 
to definitely rule out any of the models on the basis of the stellar mass distribution, its flat logarithmic slope at small masses (less prone to the above systematics, see Marchesini et al. 2009) is better matched by the WDM model. 

\begin{center}
\scalebox{0.67}[0.67]{\rotatebox{-90}{\includegraphics{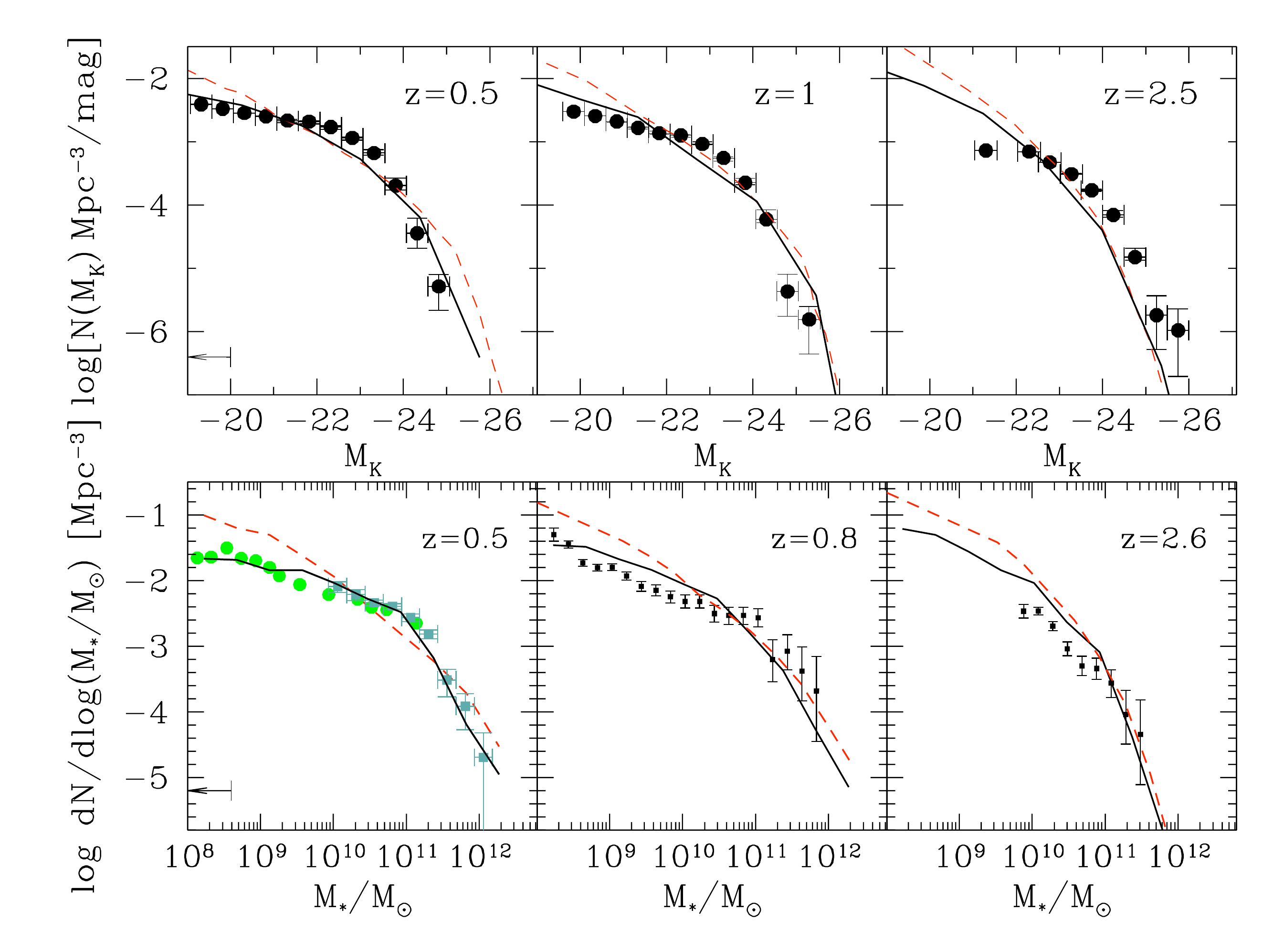}}}
\end{center}
 {\footnotesize 
Fig. 4. - Top Panel. The evolution of the K-band luminosity functions in the WDM cosmoligy (solid line) is compared with data from Cirasuolo et al. (2010). Dashed line refer to the standard CDM case. \\
Bottom Panel The evolution of the stellar mass function in the WDM cosmology (solid line) is compared with the standard CDM case (dashed); 
data in the leftmost panel are from Drory et al. (2004, squares) and Fontana et al. (2006, circles); data in all remaining panels are from Santini et al. (2012).  \\
In both the upper and the lower left panels, the arrows show the range of magnitudes and stellar masses corresponding to the free streaming mass; 
the dispersion characterizing the above relations are related to the stochastic nature of the merging trees. }
\vspace{0.2cm }

Thus, also in this case both effects (flattening of the faint end slope and sharpening of the cutoff at the bright end) are effective at low redshift,
providing a substantial improvement of the fit to the observations. At higher redshifts, the flattening at the faint end remains an approximatively constant feature, being related to the smaller number of low mass DM halo collapsed in WDM cosmology; 
although models still slightly overestimates the number of small-mass galaxies (with $M\leq 10^{10}\,M_{\odot}$) at $z\geq 0.8$, the agreement is appreciably improved by the adoption of a WDM spectrum, a result difficult to achieve adopting different feedback or star formation recipes within the CDM framework (see the comparison with other SAM in CDM cosmology in Fontanot et al. 2009; Santini et al. 2012). 

Interestingly, the effect of adopting a WDM spectrum on the bright end appears at redshift $0.5\leq z\lesssim 0.8$; this indicates that it is indeed related to the later phase of stellar mass growth in massive galaxies, that associated with the accretion of small lumps onto a central dominant galaxy. In fact, earlier comparison of different SAM predictions with observed K-band luminosity functions (see Henriques et al. 2011) have shown that the CDM over prediction of massive galaxies is caused by excessive mass growth in this redshift range. 

To investigate this point in detail, we first show in Fig. 5 (left panel) 
the number of satellites (with mass $M\geq 2\,10^9\,h^{-1}$ M$_{\odot}$) predicted by our SAM in the WDM case (normalized to that predicted in the CDM case) as a function of the host DM halo mass; we also compare our results with N-body simulations 
with the same lower mass limit for satellites (Bode et al. 2001). Note that the distribution of the satellite suppression factor in Fig. 5 (left panel) is close to that computed from N-body simulations, although the comparison can only be approximate since the distribution of the suppression factor from the N-body has been compute by Bode et al. (2001) for WDM particles with mass $m_X=0.35$ keV, while the  recent simulations by Kamada (2011) and Polisensky \& Ricotti (2011) - both with WDM particle mass $m_X=1$ keV-  have been computed only for Milky Way-size haloes. A suppression factor $\approx 5$ for Milky Way-size has been found also for $m_X=0.6-1.2$ keV by Colin et al. (2000; see also Knebe et al. 2002).  

\begin{center}
\vspace{-0.1cm}
\scalebox{0.5}[0.5]{\rotatebox{00}{\includegraphics{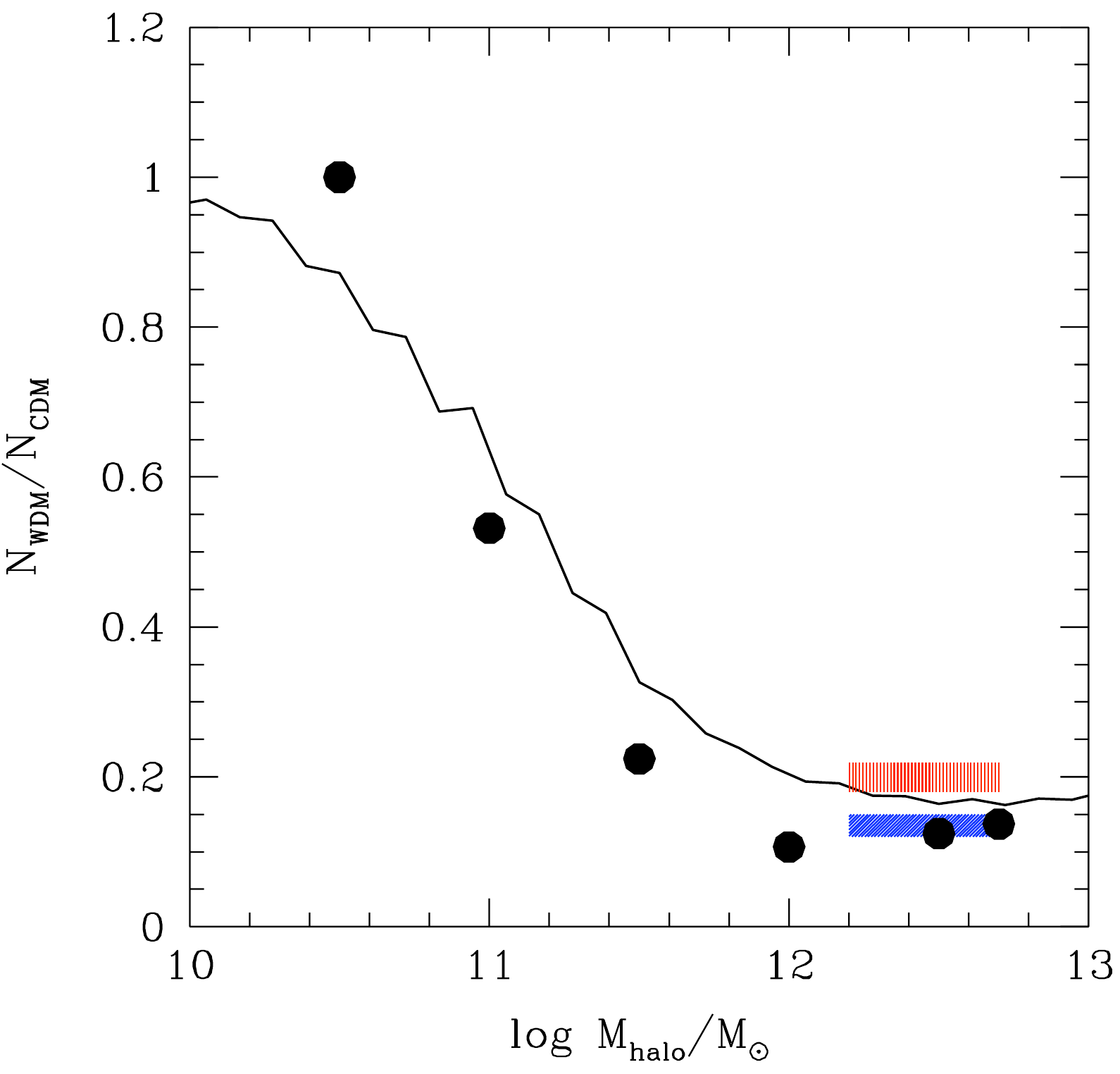}}}
\scalebox{0.6}[0.6]{\rotatebox{0}{\includegraphics{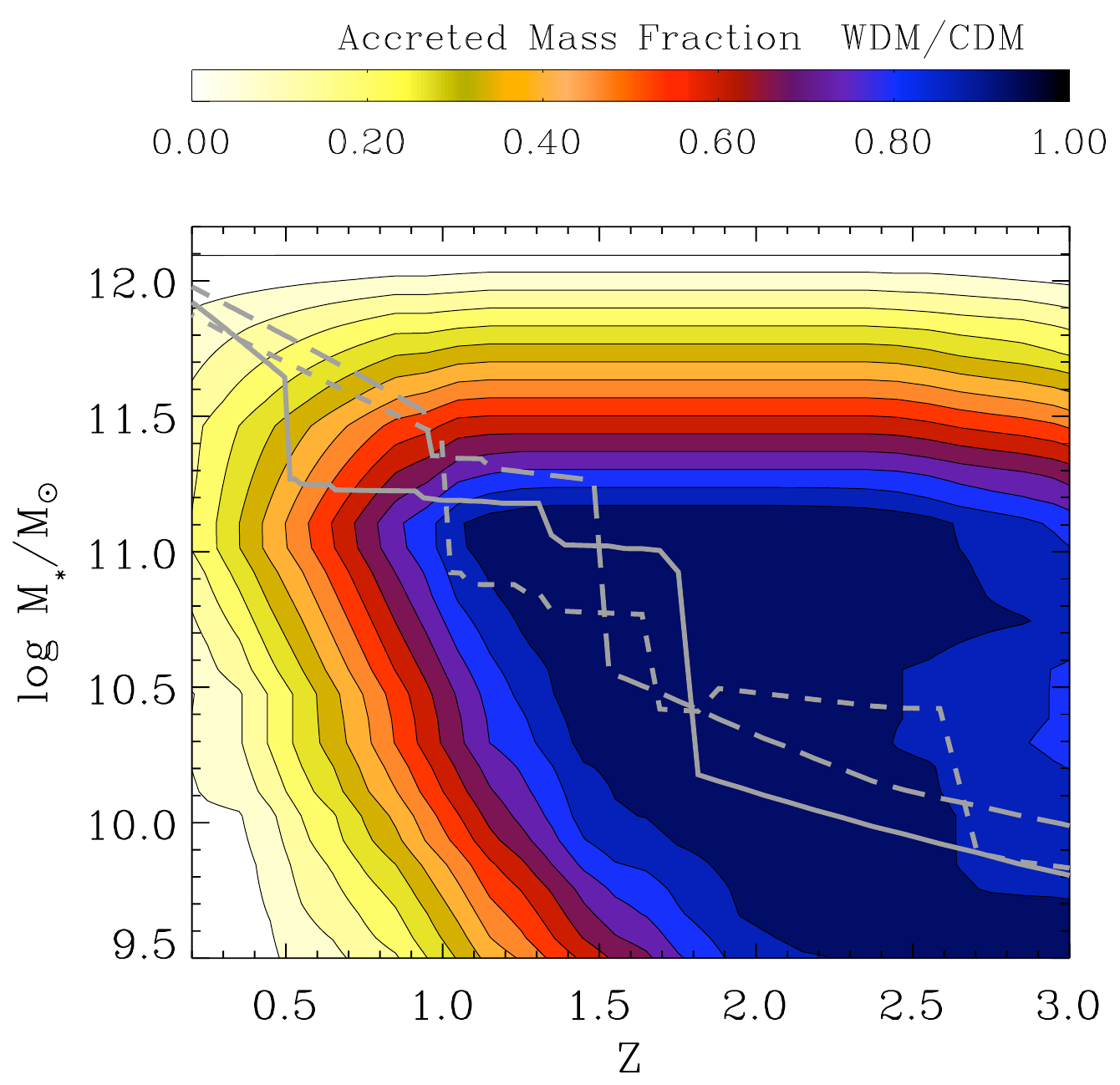}}}
\end{center}
\vspace{0.1cm }
 {\footnotesize 
Fig. 5. -  Left Panel: The predicted ratio of the number of satellite galaxies in the WDM cosmology over the corresponding CDM value is plotted against the mass of the host DM halo (solid line). We only consider satellites with $M\geq 2\,10^9\,h^{-1}$ M$_{\odot}$, in order to compare with the corresponding distribution obtained (for WDM particles with mass $m_X=0.35$ keV)
from N-body simulations by Bode et al. (2001, dots). We also represent a rendition (corresponding to our adopted lower limit on the satellite mass) of the recent results from N-body simulations of 
Milky Way-sized haloes by Kodama  (2011, upper  hatched region) and Polisensky \& Ricotti (2011, lower hatched region), both obtained with $m_X=1$ keV. \\
Right Panel: The figure illustrates the fraction of stellar mass accreted (from the inclusion of all satellites) at a given redshift (x-axis) by galaxies of given stellar mass (y-axis): the color code represents the ratio of such quantity between the WDM and the CDM case. We also show three paths representative of the growth history of massive galaxies. }
\vspace{0.3cm}

Thus, our SAM seems to provide a reasonable description of the effect of WDM 
on the the sub-halo population.  The predicted  suppression of the satellite number in the WDM case impacts the growth of massive galaxies since: i) they are hosted in larger and larger haloes for decreasing redshifts,  ii) their merging histories is dominated by the accretion of much smaller, satellite galaxies for $z\lesssim 1$ (see Baugh 2006), and iii) their low star formation at low redshift affects their stellar mass growth only mildly. Such an impact, and why it becomes appreciable for $z\lesssim 1$, is better shown in the right panel of Fig. 5, where 
we show the fraction of stellar mass gained by accretion of satellites by galaxies with different  stellar mass found at different redshift;  the color code shows such a fraction in the WDM case normalized to that in the CDM case. Note that at high redshifts, when galaxies have small stellar masses, the accretion of satellites in WDM cosmology is similar to that predicted in CDM scenarios, since at high redshifts 
merging between comparable clumps is more frequent, the number of satellites is however small, and thus the impact of minor merging on the accretion of stellar mass is small.  However, as cosmic time proceeds (for $z\lesssim 1$), satellite galaxies progressively  accumulate in larger and larger DM haloes, while the merging histories of massive objects are dominated by the infall of small mass clumps; at this stage,  the lower number of satellites predicted in the WDM scenario effectively suppresses the accreted stellar mass with  respect to the CDM case. To better illustrate the epoch when such a suppression is effective, we also show in Fig. 5 (right panel)  
the growth history of three typical massive galaxies (with mass $M_*\approx 10^{12}\,M_{\odot}$ at $z=0$); note how all such paths cross the transition region (where the suppression of accretion in the WDM becomes appreciable, see the color scale) at redshifts $z\approx 0.8-1$. This explains why for $z\lesssim 0.8$ the population of massive galaxies shows a slower increase in the WM D case compared to CDM, as shown by Fig. 4. 

\section{Discussion: Robustness with Respect to Theoretical Modelling}
We turn now to discuss the robustness of our results and conclusions when the uncertainties in the theoretical modelling of the abundance of low-mass DM haloes in WDM are considered. 
In the previous section, we adopted a straightforward extended Press \& Schechter approach to compute the abundance of haloes for a WDM power spectrum, and to generate the 
DM merging trees through a Monte Carlo SAM. However, at present, uncertainties exist in the determination of the abundance of halos with masses smaller than the free streaming mass $M_{fs}=(4\,\pi/3)\,(r_{fs}/2)^3\,\overline{\rho}_X$ (where $r_{fs}$ is given by eq. 2 and $\overline {\rho}_X$ is the average WDM density). 
There is some debate as to whether haloes with such masses can form at all  (Bode et al. 2001, Wang \& White 2007); 
recent results from numerical simulations (see Wang \& White 2007; Zavala et al. 2009)
show that, when the effects of spurious clumps associated with the limited resolution of simulations are accounted for, 
a strong suppression in the number of haloes with $M\leq M_{fs}$ occurs  (see also Schneider et al. 2011). 

Thus, it is important to explore the maximal effect that such an uncertainty can have on our results.
To this aim, we assumed a cut off in the halo abundance for $M\leq M_{fs}$. Following the approach in Smith \& Markovic (2011), 
we adopted an {\it ad hoc} suppression factor in the form of an error function, so that the mass function $dn/dlog M$ in WDM becomes
\begin{equation}
{d{\tilde n} \over d log M}={1\over 2}\,
\Big\{
1+erf\Big[{log_{10}(M/M_{fs})\over \sigma_{log M}}\Big]
\Big\}\,{dn \over d log M}~,
\end{equation}
where $\sigma_{log M}$ controls the logarithmic width of the step (see left panel of fig. 6). 
Thus, we run our Monte Carlo computations and the whole semi-analytic model after suppressing the 
abundance of DM haloes with masses $M\leq M_{fs}$ after the above equation, assuming $\sigma_{log M}=0.5$  (as in Smith \& Markovic 2011) or $\sigma_{log M}=0.1$ (corresponding
in practice to a complete suppression of the abundance of DM haloes with mass below the free-streaming mass). 
To implement such a suppression  in our Monte Carlo realizations of merging trees, we first run a Monte Carlo 
simulation with probabilities given by the EPS theory with the WDM power spectrum (see sect. 2.1 and. 2.2). This generates a set of merging 
trees where - at each time level - the average number of generated DM haloes as a function of  mass reproduces the standard Press \& Schechter mass 
distribution $dn / d log M$ at the corresponding cosmic time;  then, before running the full SAM on such merging trees, 
we suppress the number of low-mass halos in the trees by randomly eliminating haloes of mass $M$ with a suppression probability  
$f(M)=1-(d{\tilde n} / d log M)/(dn/ d log M)$. By construction, we end up with DM merging trees where the abundance of DM halos follows eq. (5).
Note that, when  $\sigma_{log M}=0.1$ is adopted in eq. (5), $f(M)\approx 1$ for all haloes with 
mass $M<M_{fs}$, leading in practice to a complete removal of sub-$M_{fs}$ DM haloes from our merging trees.

\begin{center}
\vspace{-0.5cm}
\scalebox{0.67}[0.67]{\rotatebox{-90}{\includegraphics{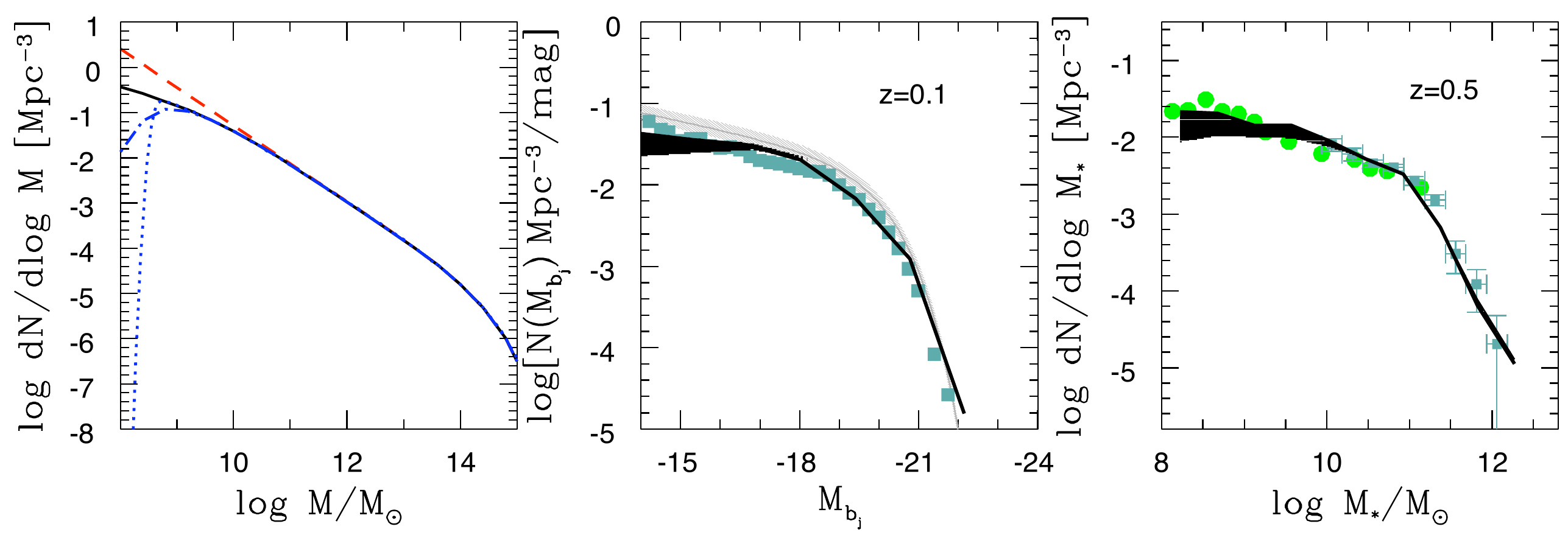}}}
\end{center}
\vspace{-0.1cm }
 {\footnotesize 
Fig. 6. - . Left panel: the halo mass function in our fiducial WDM case (solid line) is compared with that obtained after 
the adoption of a cut off (eq. 5) at the free streaming scale corresponding to $\sigma_{log M}=0.5$ (dot-dashed line) and to 
$\sigma_{log M}=0.1$ (dotted line); the dashed line refers to the CDM halo mass function. \\
Central Panel: The uncertainty in the predicted $b_j$-band luminosity function (the black filled area) associated with the uncertainty in the
abundance of low-mass haloes in WDM: the upper envelope of the black region corresponds to the WDM predictions in Sect. 3, while 
the lower envelope represents the WDM predictions after the adoption of the sharper cut off in the halo mass distribution with $\sigma_{log M}=0.02$ 
(see eq. 5 and the corresponding dotted line in the left panel); data points as in left panel of fig. 2 
Right Panel: same as central panel, but for the stellar mass distribution; data points as in the bottom-left panel of fig. 4.
\vspace{0.2cm}}

The  effects on the low-redshift luminosity function and on the stellar mass 
functions at $z=0.5$ are shown in the fig. 6. 
The overall result is a further flattening of the luminosity/stellar mass functions, caused by the lack of low-mass DM haloes. 
However, when passing from the abundance of DM halos to galaxy counting (in luminosity or stellar mass bins),  
the effect of suppressing the 
abundance of sub-$M_{fs}$ haloes is smeared out over a range of luminosities/stellar masses
due the concurrence of two effects. On the one hand, the effect of  suppressing the
abundance of sub-$M_{fs}$ haloes is not concentrated on lowest-luminosity objects,
since - even when sub-$M_{fs}$ haloes are allowed to form at some redshifts - only a fraction of 
the galaxies formed in them remain as an isolated clump, the remaining fraction being included into 
a larger galaxy (with a larger luminosity/stellar mass) at some lower redshifts. 
On the other hand, the stochastic nature of the star formation histories in SAMs results 
a) in a relevant fraction of the faintest galaxies to form in halos with masses $M>M_{fs}$  (whose abundance is not suppressed), and 
b) in a relevant fraction of sub-$M_{fs}$ halos (when they are allowed to form) to host galaxies with 
 extremely low stellar mass and star formation rate, so that they contribute only 
to luminosity and stellar masses well below the observational threshold. 

Thus, the theoretical uncertainty on the abundance of DM halos with $M<M_{fs}$ results in the uncertainty regions 
for the predicted luminosity and stellar mass functions  shown in fig. 6.  A similar degree of uncertainty is obtained for 
all the luminosity functions shown in Sect. 4, illustrating how our basic conclusions are robust with respect to 
present uncertainties in the modelling of sub-$M_{fs}$ WDM haloes. 

\section{Summary and Conclusions}

In this paper we investigated the effect of assuming a WDM power spectrum on galaxy formation using a SAM, leaving unchanged the baryonic mechanisms already implemented in  our previous CDM  model (in particular the stellar and the AGN feedback); this allows to single out the effects of changing the DM spectrum with the same baryon physics. We focussed on 
the statistical distribution of luminosity and stellar masses, leaving to next paper a full study of the effects of WDM on other properties of the galaxy population (e.g., chemical abundances, ages of stellar populations, effects of the environment). Within the present study, 
we find that adopting power spectrum corresponding to WDM particles with mass $m_X\approx 1$ keV 
may contribute to solve two major problems of CDM galaxy formation scenarios, namely, the excess  of predicted faint (low mass) galaxies at low and especially at high redshifts,  and the excess of bright (massive) galaxies at low redshifts. In fact, adopting a WDM spectrum on the one hand results in a smaller number of collapsed low-mass haloes; on the other hand it results into a smaller number of satellite galaxies accumulating in massive haloes at low redshifts, thus suppressing the accretion of small lumps on the central, massive galaxies for $z\lesssim 0.8$. 
Such conclusions are robust with respect to the uncertainties in the modelling of the WDM halo abundance around the free streaming mass scale, as discussed in Sect. 4.

Our results on the luminosity and stellar mass distributions of galaxies concur to indicate that particles with mass $\approx 1$ keV  constitute interesting candidates for DM; such an indication adds to the existing astrophysical evidences from (see references in the Introduction): a) the observed galaxy structures at small scales ($\leq 50$ kpc); b) the cored profiles of galaxies; c) the number of sub haloes and satellites in the haloes of massive galaxies. 

Directly testing whether DM is constituted by such particles, is a realistic possibility for the near future. A cosmological direct probe of  WDM could be provided by cosmic shear (weak gravitational lensing), which has the advantage of being weakly dependent on baryonic physics. First results in the simple relic scenario indicate that future lensing surveys like EUCLID could see WDM signals for $m_X\gtrsim 2$ keV (Smith \& Markovic 2011), close to the mass adopted here. 

If the WDM is constituted by sterile neutrinos, as advocated to account for the non-zero active neutrino mass (for a review see Boyarsky, Ruchayskiy \& Shaposhnikov 2009), astrophysical bounds on its mass can be placed because it would decay in a standard neutrino plus a photon with energy of the order of the keV, thus producing narrow line emission in X-rays. Claims for detection of lines hardly identifiable with known transitions have been published using both Chandra and Suzaku observations of dwarf galaxies and the Galactic centre (Loewenstein \& Kuseno 2010, Prokhorov \& Silk 2010). Both the claimed lines are at energies where the effective area is strongly reducing (8.7 keV Suzaku) or near the Gold M edge feature of the Chandra mirror (2.51 keV). In the case WDM is constituted by sterile neutrinos direct detection is also possible through triple beta decay experiments like MARE and KATRIN (see de Vega et al. 2011). 

As for the case of thermal WDM, the favourite candidate is the gravitino. It emerges as a natural candidates for the lightest supersymmetric (LSP) particle in  gauge-mediated supersymmetry breaking models (see Dine, Nelson, Nir, Shirman 1996) with conserved R parity. Contrary to the gravity-mediated supersymmetry breaking case, where the LSP (typically the neutralino) has masses exceeding 
the 10 TeV scale, the gravitino mass can range from a fraction of eV up to O(GeV). However, if it constitutes the LSP with mass $m_X\approx 1$ keV,  a large value $g_*\approx 10^3$ for the degrees of freedom would be needed at the time of its decoupling to comply with the observed matter density $\Omega_M\,h^2\approx 0.12\approx (100/g_*)\,(m_X/{\rm 1 keV})$ (Pagels \& Primack 1982); this is larger than $g_*\sim 10^2$  expected in minimal supersymmetric standard model (MSSM). Of course, scenarios with $g_*$ exceeding the MSSM expectations  are conceivable, as well as scenarios where entropy-producing processes could dilute the gravitino abundance (Fujiii and Yanagida 2002), or where vanishing gauge coupling at high temperature suppresses the gravitino production (Buchmuller, Hamaguchi \& Ratz 2003; see also Lemoine, Moultaka, Jedamzik 2006).  A gravitino mass $m_X\approx 1$ keV effective 
to provide a viable WDM scenario for galaxy formation would correspond to a relatively low energy scale for 
supersymmetry breaking $\Lambda\approx (m_X\,M_p)^{1/2}\sim 10^6$ GeV (here $M_p$ is the reduced Plank mass). 
While its extremely weak coupling with matter 
($\propto \Lambda^{-2}$, see Lee \& Wu, 1999) makes gravitino DM inaccessible to direct searches, upcoming data from colliders may allow for signals from light gravitinos with mass in the eV to MeV range (see Feng, Kamionkowski \& Lee 2010). Light gravitinos would be produced primarily in the decays of the next-to-lightest supersymmetric particle, resulting in a variety of signals, including di-photons, delayed and nonpointing photons, kinked charged tracks, and heavy metastable charged particles. If light gravitino constitutes WDM, the 7 TeV LHC with 1  fb$^{-1}$ could see evidence for hundreds of light-gravitino events. 

Our results  have straightforward implications concerning the growth of Supermassive Black Holes and the evolution of AGNs. 
In cosmological galaxy formation theories, these are tightly related to the evolution of their host galaxies, as indicated also by the 
observed correlation between the mass of Supermassive Black Holes and global properties of their host galaxies, like the stellar mass  or the velocity dispersion   (Kormendy \& Richstone 1995; Magorrian et al. 1998; Gebhardt et al. 2000; Ferrarese \& Merritt 2000). As recalled in the Introduction (see references therein),  the over-predictions of low-mass objects in the CDM scenario concerns also the AGN population; assuming a WDM power spectrum could substantially alleviate the problem, analogously to what we showed to happen for the galaxy population. Note that, contrary to what we derived for low redshift massive galaxies, we do not expect bright AGNs to be affected by assuming a WDM spectrum. This is because such objects are thought to be powered by gas accretion triggered by galaxy encounters with partners of comparable mass (extremely rare at low redshift), only mildly affected by the abundance of low mass satellites. We shall investigate this points in a next paper.

We thank M. Castellano, A. Fontana, P. Santini, and the referee for helpful comments and suggestions. We acknowledge financial support from the agreement ASI-INAF 1/009/10/0.\\

\end{document}